\documentclass[10pt]{iopart}
\expandafter\let\csname equation*\endcsname\relax
\expandafter\let\csname endequation*\endcsname\relax
\usepackage{amsmath}
\usepackage{amssymb}
\usepackage{tabularx}
\usepackage[switch]{lineno}
\usepackage{cite}
\usepackage{mathtools}
\usepackage{times}
\usepackage[usenames,dvipsnames]{color}
\usepackage{graphicx}
\usepackage{epsfig}
\usepackage{epstopdf}
\usepackage{dcolumn}
\usepackage{bm}
\usepackage{array}
\usepackage{lipsum}
\usepackage{simplewick}
\usepackage{romannum}
\usepackage{booktabs}
\usepackage{siunitx}
\usepackage{float}
\usepackage{enumitem}
\usepackage{url}
\usepackage[utf8]{inputenc}
\usepackage[T1]{fontenc}
\usepackage{mathptmx}
\usepackage[colorlinks,citecolor=blue,urlcolor=blue,bookmarks=false,hypertexnames=true]{hyperref}

\begin{document}
\title{Spectroscopy of the $5s5p$ $ ^3 P_0 \rightarrow 5s5d$ $ ^3 D_1 $  transition of strontium using laser cooled atoms}

\author{Kushal Patel$^1$, Palki Gakkhar$^2$, Korak Biswas$^1$, S. Sagar Maurya$^1$, Pranab Dutta$^1$, Vishal Lal$^1$, B. K. Mani$^2$, and Umakant D. Rapol$^1$}

\address{$^1$ Department of Physics, Indian Institute of Science Education and Research, Pune 411008, Maharashtra, India}
\address{$^2$ Department of Physics, Indian Institute of Technology Delhi, New Delhi 110016, India}
\ead{umakant.rapol@iiserpune.ac.in}

\date{\today}
\begin{abstract}
This article presents spectroscopy results of the $5s5p{\;^3}P_0 \rightarrow 5s5d{\;^3}D_1$ transition in all isotopes of laser cooled Sr atoms and the utility of this transition for repumping application. By employing the $5s5p{\;^{3} P_{0}} \rightarrow 5s5d{\;^3}D_1 $ (483 nm) transition in combination with the excitation of $5s5p{\;^3}P_2 \rightarrow 5s6s{\;^3}S_1$ (707 nm) transition, we observe a significant increase ($\sim$ 13 fold) in the steady state number of atoms in the magneto-optic trap (MOT). This enhancement is attributed to the efficient repumping of Sr atoms that have decayed into the dark $5s5p{\;^3}P_2$ state by returning them to the ground state $5s^2{\;^1}S_0$ without any loss into the other states. The absolute transition frequencies were measured with an absolute accuracy of 30 MHz. To support our measurements, we performed Fock-space relativistic coupled-cluster calculations of the relevant parameters in Sr. To further increase the accuracy of the calculated properties, corrections from the Breit, QED and perturbative triples were also included. The calculated branching ratio for the repumping state confirms the significantly increased population in the ${^3}P_1$ state. Thereby, leading to an increase of population of atoms trapped due to the enhanced repumping. Our calculated hyperfine-splitting energies are in excellent agreement	with the measured values. Moreover, our calculated isotope shifts in the transition frequencies are in good agreement with our measured values.

\end{abstract}

\submitto{\jpb}
\maketitle

\ioptwocol

\section{Introduction}
Laser-cooled atoms are utilized in a broad range of scientific and technological applications -- ranging from exploring fundamental physics to developing cutting-edge quantum technology \cite{metcalf, brierley} based devices. In a two-electron system like rare earth alkaline atoms, singlet and triplet energy-level manifold emerges. This give rise to narrow linewidth transitions suitable for optical frequency standards \cite{ludlow}, and other exotic applications such as atomic gravimeters \cite{grotti,stray}, interferometers \cite{avinadav} , superradiant lasers \cite{norcia}, quantum simulations \cite{schine}, and quantum computing \cite{saffman,jenkins}. Statistical sensitivity in precision experiments is directly influenced by both the no of atoms present in the cold atomic sample and the time duration of the cold sample's preparation. Therefore, a fast-loading magneto-optical trap (MOT) with high atom numbers becomes desirable.

The main cooling and trapping transition in alkaline earth atoms is not fully closed due to the decay of  atoms into a $5s5p{\;^3}P_2$ metastable state. Due to the long lifetime of this metastable state, atoms leave the cooling cycle, resulting in a significant loss of atoms from the MOT. However, the introduction of repumping lasers can help mitigate these losses. Another approach to address this issue involves the preparation of cooled atoms in a reservoir state that remains unaffected by these losses. In case of strontium atoms, when the repumping lasers are absent, the atoms gradually accumulate in a magnetically trapped metastable state ${\;^3}P_2$ during the operation of MOT, which can further be used in the experiment. The drawback of this approach is that it increases the experimental cycle time \cite{nagel}.

\begin{figure}[!h]
	\includegraphics[scale=0.4]{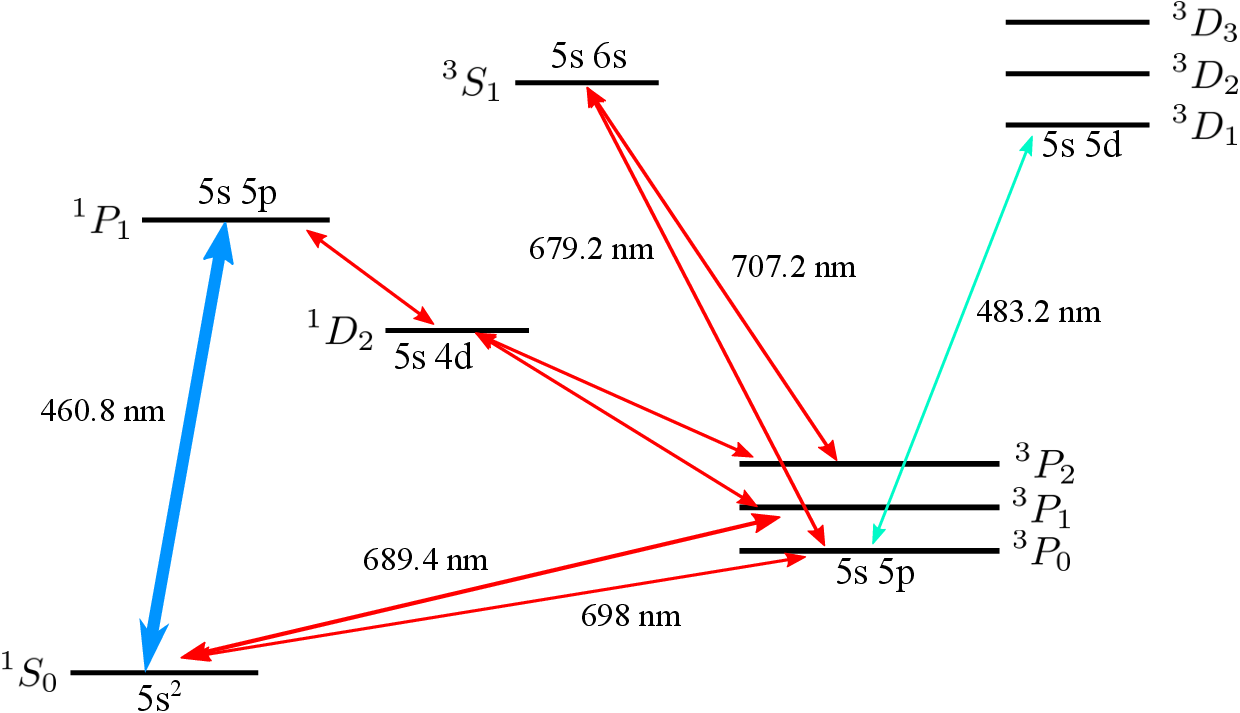}
	\caption [] { \label{fig:fig1} Schematic illustration of  the low-lying energy levels of bosonic
		strontium with wavelengths taken from  J. Sansonetti and G. Nave \cite{sanso}. The transitions at 461 and 689 nm 
		are used for MOTs, the transitions at 679 and 707 nm are the repumping 
		transitions. In this study atoms in the metastable state $5s5p$ $ ^3P_2$ 
		are repumped using the 707 nm, and $5s5p$ $^3P_0 \rightarrow 5s5d $ $^3D_1$ 
		transitions at 483 nm.}
\end{figure}

Conventionally, to repump atoms from $5s5p{\;^3}P_2$ state dual repumper lasers operating at 707 nm ($5s5p{\;^3}P_2 \rightarrow 5s6s{\;^3}S_1 $) and 679 nm ($5s5p{\;^3}P_0 \rightarrow 5s6s{\;^3}S_1 $) are used \cite{takamo} [see Fig. {\ref{fig:fig1}}]. In this study, we demonstrate a novel dual-repumping scheme, utilizing a 483 nm laser to target the $5s5p{\;^3}P_0 \rightarrow 5s5d{\;^3}D_1 $ transition, and a 707 nm laser for the driving $5s5p{\;^3}P_2 \rightarrow 5s6s{\;^3}S_1$ transition. The 483 nm  repumping laser can also be used in combination with the other single repumping schemes, such as 481 nm ($5s5p{\;^3}P_2 \rightarrow 5p^{2}{\;^{3} P_{2}}$), 496 nm ($5s5p{\;^3}P_2 \rightarrow 5s5d{\;^3}D_2$) and 405 nm ($5s5p{\;^3}P_2 \rightarrow 5s6d{\;^3}D_2$) to prevent losses in a $^3P_0$ state thereby, enhancing the number of atoms in the MOT \cite{stellmer,hu,moriya}. This transition can also be used for the clock spectroscopy to probe the number of atoms in a long-lived clock state, by coupling a $5s5p{\;^3}P_0$ state to the ground state using a 483 nm laser.

In this work, through experimental measurements and theoretical calculations, we present the branching ratio, transition linewidth, hyperfine splitting constants, and isotope shift of the $5s5p{\;^3}P_0 \rightarrow 5s5d{\;^3}D_1$ transition. We also study the dependence of atom number enhancement in the MOT by resonantly driving the $5s5p{\;^3}P_0 \rightarrow 5s5d{\;^3}D_1$ transition at 483 nm in addition to $5s5p{\;^3}P_2 \rightarrow 5s6s{\;^3}S_1$ transition at 707 nm for all the stable isotopes of strontium. We compare the performance of 483 nm repumper with 679 nm which is conventionally used as a repumper. Furthermore, in this work we measure the absolute frequency of the 483 nm transition in $^{88}$Sr, $^{87}$Sr, $^{86}$Sr, and $^{84}$Sr isotopes with an absolute accuracy of 30 MHz, and isotope shift is compared with our calculated data.

\section{Experimental and Theoretical Details}

In the experiment, atoms are loaded into a three-dimensional magneto-optical trap 
from a Zeeman slower \cite{joffe,phillips1998nobel}. A blue MOT is created using 
three retro-reflected beams of 461 nm each, with an intensity of 10 mW/cm$^{2}$, 
a detuning of -45 MHz $(\sim 1.5$ $\Gamma)$, and a beam diameter ($1/e^{2}$) of 12 mm. An axial magnetic 
field gradient of 50 G/cm is maintained throughout the experiment. The blue light 
is produced using a second harmonic generation (SHG) cavity and an external-cavity 
diode laser (ECDL). A portion of the output light from the SHG cavity is used for 
injection-lock the high-power Nichia 460 nm laser diode [NDB 4916]. The Zeeman 
slower light is obtained from the SHG output, while the light for the MOT beams 
is derived from the injection-locked output. The repumping light at 707 nm is 
generated using a home-built external-cavity diode laser (ECDL), and the 483 
nm light is generated using a commercial [Toptica DL pro] ECDL with output 
powers of 25 mW and 20 mW respectively. More details about the experimental 
set-up can be found in references \cite{biswas2023machine,biswas2023electromagnetically}.

The frequency of the laser (483 nm) used for spectroscopy is continuously monitored using a commercial wavemeter [HighFinesse WSU/30], which is regularly calibrated with a 689 nm laser locked to the Sr atom. The 689 nm laser is locked to a temperature stabilized, in-vacuum high-finesse optical cavity, exhibiting a measured drift rate of less than 200 Hz per hour. The absolute frequency is established through the spectroscopy of the $5s5p$ $ ^3P_1$ state in strontium - the frequency of which is  known to within 10 kHz \cite{ferr}. The wavemeter is specified to have an absolute accuracy of 30 MHz and a measurement resolution of 1 MHz.
\clearpage

To support our experimental data, we performed Fock-space relativistic coupled-cluster (FSRCC) calculations of the relevant experimental properties for the repumping transition demonstrated in this work. The relativistic coupled-cluster theory is one of the most accurate many-body theories for the calculation of atomic properties. In FSRCC theory, within the coupled-cluster singles and doubles approximation, the many-electron wavefunction for a {\em two-valence} electron atom or ion, like atomic Sr which is the interest of the present work,  is expressed as

\begin{equation}
	|\Psi_{vw}\rangle = e^{(T_1 + T_2)} \left[ 1 + S_1 + S_2 + 
	\frac{1}{2} \left( S_1^2 + S_2^2 \right) + 
	R_2 \right ]|\Phi_{vw}\rangle,
	\label{2v}
\end{equation}
\\
where $vw\ldots$ represent the valence orbitals and $|\Phi_{vw}\rangle$ is the Dirac-Fock reference state. In Fock-space formalism, it is obtained by adding two electrons to the Dirac-Fock state for closed-shell configuration, $a^\dagger_wa^\dagger_v |\Phi_0\rangle$. The excitation operators, $T$, $S$ and $R$ are referred to as the coupled-cluster operators for {\em closed-shell}, {\em one-valence} and {\em two-valence} sectors, respectively, and expressed in terms of the creation and annihilation operators \cite{mani2017, kumar2021, kumar2022}.
The many-electron wavefunction $|\Psi_{vw}\rangle$,  in Eq. (\ref{2v}), is obtained from the 
solution of the following eigenvalue equation

\begin{equation}
	H^{\rm DCB}|\Psi_{vw} \rangle = E_{vw} |\Psi_{vw} \rangle,
	\label{hdc_2v}
\end{equation}
\\
where $E_{vw}$ is the exact energy of the system. And, $H^{\rm DCB}$ is the Dirac-Coulomb-Breit no-virtual-pair Hamiltonian, expressed as

\begin{equation}
	\begin{aligned}
		H^{\rm DCB}  =  \sum_{i=1}^N \left [c\bm{\alpha}_i \cdot
		\mathbf{p}_i + (\beta_i -1)c^2 - V_{N}(r_i) \right ]\\
		\shoveleft{}+ \sum_{i<j}\left [ \frac{1}{r_{ij}}  + g^{\rm B}(r_{ij}) \right ].
	\end{aligned}
\end{equation}
\\
Here, $\bm{\alpha}$ and $\beta$ are the Dirac matrices, and the last two terms, $1/r_{ij} $ and $g^{\rm B}(r_{ij})$, represent the Coulomb and Breit interactions, respectively. It is to be however mentioned that, in the actual calculations, we have also incorporated the contributions from the quantum electrodynamical (QED) corrections such as the self-energy and vacuum polarization. To improve the accuracy of our computed properties further, we have also incorporated the triples excitations in coupled-cluster perturbatively. More details about the implementation of these interactions in our calculations and FSRCC theories and corresponding technical details, in general, can be found in our previous works \cite{mani2017, kumar2021, kumar2022}

After we get the wavefunction in Eq. (\ref{2v}) from FSRCC calculations, we can use it for the properties calculations \cite{mani2017, kumar2021, kumar2022}. In the present work, to support our experimental findings related to the demonstrated repumping scheme, we have calculated the excitation energy, magnetic dipole and electric quadrupole hyperfine constants and  energy splitting for the  repumping state $5s5d{\;^3}D_1$. Further, we have calculated the transition rates for $5s5p{\;^3}P_{0,1,2} \rightarrow 5s5d{\;^3}D_1$ and $5s5p{\;^1}P_1 \rightarrow 5s5d{\;^3}D_{1}$ transitions. These transition rates are used to calculate the transition linewidth and branching ratios associated with repumping transition, which are discussed later. The upper bound in the uncertainty of our calculated properties using FSRCC is $\approx$ 4.5\% \cite{kumar2021}. In addition, to support our isotope shift measurements, we have calculated the isotope shifts in transition frequencies using multiconfiguration Dirac-Hartree-Fock (MCDHF) theory, results from which are provided and discussed later.

\section{MOT REPUMPING SCHEME}

%%%%%%%%%%%%%%%%%%%%% figure %%%%%%%%%%%%%%%%%%%%%%%%%%%%%%%%%%%%%%%%%%
%%%%%%%%%%%%%%%%%%%%%%%%%%%%%%%%%%%%%%%%%%%%%%%%%%%%%%%%%%%%%%%%%%%%%%%%

In Figure 1, the energy levels and transitions relevant to the laser cooling and trapping of strontium are illustrated. Atoms initially in the $5s^{2}$ $ {^{1} S_{0}}$ ground state are driven to the $5s5p $ $^1P_1$ state using a 461 nm laser. The $5s5p $ $^1P_1$ state has a decay probability of 1 in 50,000 to the $5s4d$ $ ^1D_2$ state, which further decays to the $5s5p$ $ ^3P_{1,2}$ with a branching ratio of 2:1. Meanwhile, atoms in the $5s5p$ $ ^3P_2$ state are excited to the $5s6s$ $ ^3S_1$ state using a 707 nm laser, which subsequently decay to the $5s5p $ $^3P_{0,1,2}$ states with a branching ratio of approximately 5:3:1. The atoms in the $5s5p$ $ ^3P_0$ state are excited to the $5s5d$ $ ^3D_1 $ state using a 483 nm laser, which decays to the $5s5p$ $ ^3P_{0,1,2}$ states with a branching ratio of approximately 56:41:3 \cite{stellmer}, ultimately populating only the $5s5p$ $ ^3P_1$ state. Atoms in the short-lived $5s5p$ $ ^3P_1$ state decay back to the ground state thereby again making them participate in the cooling cycle, hence inhibiting the loss in the atom number

To get more insight into the repumping transition and associated properties, using the transition probabilities listed in Table \ref{prop_tab}, we calculated the branching ratio and transition linewidth for the repumping transition. As evident from the table, to increase the accuracies of these properties, the corrections from the Breit and QED interactions and perturbative triples are also incorporated at the level of transition probabilities. As mentioned earlier, the proposed repumping transition could also be useful for the atomic clocks experiments, and therefore the knowledge about the branching ratios of ${^3}D_1$ into ${^3}P_{0,1,2}$ and ${^1}P_{1}$ states could be crucial for minimizing the statistical uncertainty. 
The branching ratio of ${^3}D_1$ to a lower atomic state function (ASF) $|\Psi_i\rangle$ is calculated as

\begin{equation}
	\gamma_{{^3}D_1 \rightarrow |\Psi_i\rangle} = \frac{A_{{^3}D_1 \rightarrow |\Psi_i\rangle}}
	{\sum_{i}A_{{^3}D_1 \rightarrow |\Psi_i\rangle}},
\end{equation}
\\
where $A_{{^3}D_1 \rightarrow |\Psi_i\rangle}$ represents the electric dipole (E1) transition probability and $|\Psi_i\rangle$ are the dipole allowed ASFs, ${^3}P_{0,1,2}$ and ${^1}P_{1}$. The transition probability is further expressed in terms of the transition line strength, $S_{\rm E1}=|\langle {^3}D_1||E1|| \Psi_i \rangle||^{2}$, and the wavelength of the transition, as \cite{johnson}

\begin{equation}
	A_{{^3}D_1 \rightarrow |\Psi_i\rangle} = \frac{2.02613\times10^{18}}{3 \lambda^{3}} S_{\rm E1}. 
\end{equation}
\\
Here, the transition line strength is in atomic units and $\lambda$ is in angstrom. 
As evident, $S_{\rm E1}$ is square of the reduced dipole matrix elements, which is 
calculated using FSRCC theory.

\begin{figure}[h]
	\includegraphics[width= 5cm, angle=-90]{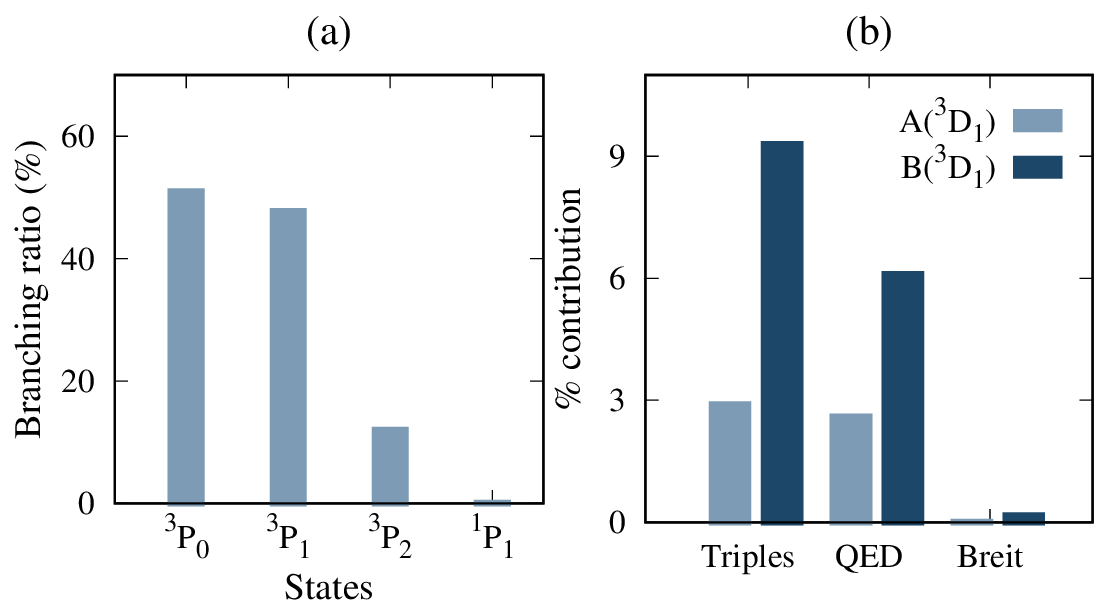}
	\caption [] {(a) Branching ratio of repumping state $5s5d{\;^3}D_1$
		in $5s5p{\;^3}P_{0,1,2}$ and $5s5p{\;^1}P_1$ states. (b) Percentage 
		contributions from the Breit, QED and perturbative triples corrections 
		to hyperfine constants.}
	\label{br_fig}
\end{figure}

In Fig. \ref{br_fig}, we have shown our data on the branching ratio for repumping state ${^3}D_1$. As discernible from the figure, the nonzero branching ratios suggest 
that electrons from ${^3}D_1$ state can make a transition to ${^3}P_{0, 1, 2}$ and ${^1}P_1$ states via E1 transitions. The probability of transition is, however, different for different states. Consistent with our proposed repumping scheme, the population in $5s5p{\;^3}P_2$ is significantly reduced. And, it is as large as $\approx$  51\% and 48\%, respectively, for ${^3}P_0$ and ${^3}P_1$ states. This is as desired, as these electrons can transition back to ground state ${^1}S_0$ via radiative and E1 transitions, respectively. The population in ${^1}P_1$ is negligible, and the reason for this could be attributed to the smaller transition probability due to closer energy with ${^3}D_1$. The transition probabilities for allowed dipole transitions are further used to calculate the transition linewidth, expressed as $2\pi \times \gamma = \sum_{i} A_{{^3}D_1 \rightarrow |\Psi_i\rangle}$. Our calculated value $\gamma$ = 10.2 MHz, is in good agreement with the value of 9.4 MHz  and 9.7 MHz obtained from previous theory\cite{porsev} and experiment \cite{andra} respectively. It is to be mentioned that, in the calculation of $\gamma$ we have used the experimental energies to increase its accuracy. The small difference in our calculated value and Ref. \cite{porsev} could be attributed to the use of slightly different many-body methods in the two calculations and the corrections from the Breit and QED incorporated in our results. From the Breit and QED contributions listed in Table \ref{prop_tab}, we find that Breit+QED contributes $\approx$ 3.2\% to the linewidth.

%%%%%%%%%%% figure%%%%%%%%%%%%%%%%%%%%%%%%%%%%%%%%%%%%%%%%%%%%%%%%%%%%%%%
%%%%%%%%%%%%%%%%%%%%%%%%%%%%%%%%%%%%%%%%%%%%%%%%%%%%%%%%%%%%%%%%%%%%%%%%%

\section{CHARACTERIZATION OF THE REPUMPING SCHEME}

To characterize the repumper performance, we compare the steady state atom number in the MOT in the presence and absence of repumping light with different schemes. For each scheme of repumping, the MOT is operated for a fixed time of 2 s and the fluorescence is recorded on a photo multiplier tube (PMT). The enhancement factor ($\eta$) in atom number is calculated by comparing the steady-state atom numbers in the MOT with and without repumping laser(s), expressed as a ratio. For different isotopes of Sr, the enhancement factor ($\eta$) is summarized in table \ref{tab:tab1}. The peak atom number in the continuously repumped blue MOT for $^{88}$Sr reaches $5\times 10^{7}$. Additionally, we perform absorption imaging using a CCD camera to determine the temperature of the cloud using the time-of-flight method. Strontium atoms in the blue MOT is at temperatures of  $\sim$ 2 mK. Furthermore, these laser-cooled strontium atoms are utilized for conducting all the isotope-shift spectroscopy.\

%%%%%%%%%%%%%%%%%%%%%%%%%%%%%%%%%%%TABLE%%%%%%%%%%%%%%%%%%%%%%%%%%%%%%%%%%%%%%%%%%
\begin{center}
	
	\begin{table}[!h]
		\caption[]{\label{tab:tab1} The enhancement of the steady-state atom 
			number in the MOT varies for different strontium isotopes 
			in various repumping scenarios. The 707 nm repumper depopulates 
			the $5s5p$ $ ^3P_2$ state, while the 483 nm repumper depopulates 
			the $5s5p$ $ ^3P_0$ state --  effectively depopulating the 
			triplet manifold via short lived $5s5p $ $^3P_1$ state. The 
			enhancement factor is normalized with respect to the steady 
			state number of atoms in the MOT in the absence of any repumping light.}
		\begin{tabular*}{\columnwidth}{@{\extracolsep{\stretch{1}}}*{4}{c}@{}} 
			\hline
			\hline
			{Isotope} & {} & {Enhancement factor} & {} \\
			
			{} & {No} & { Only 707 nm } & {707 and 483 } \\
			
			{} & {repumper} & {repumper} & {nm repumper} \\
			
			\midrule
			88& 1 & 4  & 12.8 (5)    \\
			86& 1 & 3 & 18.2 (5)  \\
			87& 1 & 5 &  18.6 (5)  \\
			84& 1 & 5 & 19.1 (5)  \\
			\hline
			\hline
		\end{tabular*}
		
	\end{table}	
\end{center}

%%%%%%%%%%%%%%%%%%%%%%%%%%%%%%%%%%%%%%%%%%%%%%%%%%%%%%%%%%%%%%%%%%%%%%%%%%%%%%

To get insight into MOT dynamics, we model the steady-state atom numbers in the MOT using a time-dependent trap loading equation. In the limit of one-body and two body losses, the trap loading equation can be written as \cite{zhang,mickelson}\\

\begin{equation}  
	\dot{N} = L- \Gamma N - {\beta}' N^{2},  
	\label{eqn5}          
\end{equation}
\\
where  N is the number of atoms, L is the loading rate of MOT, $\Gamma$ is the one-body loss rate, and $\beta^{'} = \beta / (2\sqrt{2}V)$, where $\beta$ is the two-body loss constant and V is the effective volume of the two body processes. The solution to the above differential equation (Eqn. \ref{eqn5}) can be obtained as\\
\\
\begin{equation}  
	N(t) =  \dfrac{N^{\rm ss} (1-e^{-\gamma t})}{ (1+\chi e^{-\gamma t})}    
\end{equation}\\
\\  
where $N^{\rm ss}$ is the steady-state MOT atom number, $\gamma, = \Gamma + 2 \beta^{'} N^{ss},$ represents the total loss rate and $\chi, = \frac{\beta^{'} N^{\rm ss}}{\beta^{'} N^{\rm ss}+\Gamma},$ is a measure of the relative contributions of one- and two-body loss coefficients.

\begin{figure}[!h]
	\includegraphics[scale=0.58]{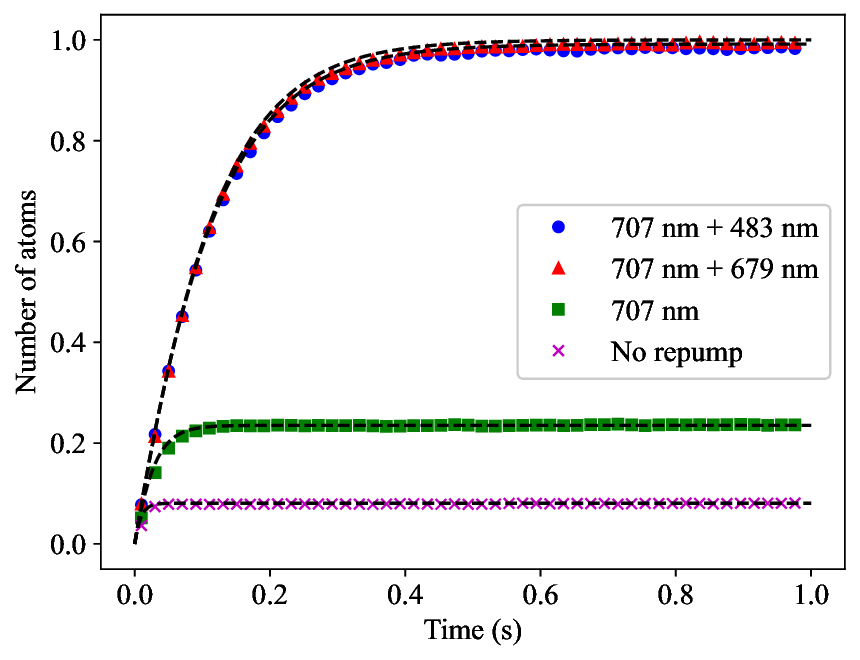}
	\caption []{ \label{fig:fig2} The normalized fluorescence of the blue 
		MOT for $^{88}$Sr for various chosen repumping schemes. 
		The black dashed curves are the fits to the Eq. (2), and the colored markers 
		represent the experimental loading curve. It is important to note 
		that the loading rates of the MOT are identical for both the 
		707 nm + 679 nm and the 707 nm + 483 nm repumping schemes.}
\end{figure}

We have used this mathematical model to fit our experimental data for three different repumping schemes, as shown in Figure \ref{fig:fig2}. From these fits we extracted the rates of one-body and two-body losses. In the case of the 483 nm scheme, the two-body fit yields $\Gamma$ = 5.16(5) $s^{-1}$ and $\beta$ = 2.06(5) $\times$ $10^{-10}$ $\text{cm}^{3}/\text{s}$. And for the 679 nm scheme, we obtain $\Gamma$ = 5.25(5) $s^{-1}$ and $\beta$ = 2.81(5) $\times$ $10^{-10}$ $\text{cm}^{3}/\text{s}$. The close values of the parameters with respect to each other suggest that both the 483 nm and 679 nm repumpers have similar MOT loading performance..

%%%%%%%%%%%%%%%%%%%%%%%%%%%%%%%%% figure 

%%%%%%%%%%%%%%%%%%%%%%%%%%%%%%%%%%%%%%%%%

\section{SPECTROSCOPY OF THE $5s5p$ $ ^3 P_0 \rightarrow 5s5d$ $ ^3 D_1 $ TRANSITION}

\begin{figure}[!h]
	\includegraphics[scale=0.55]{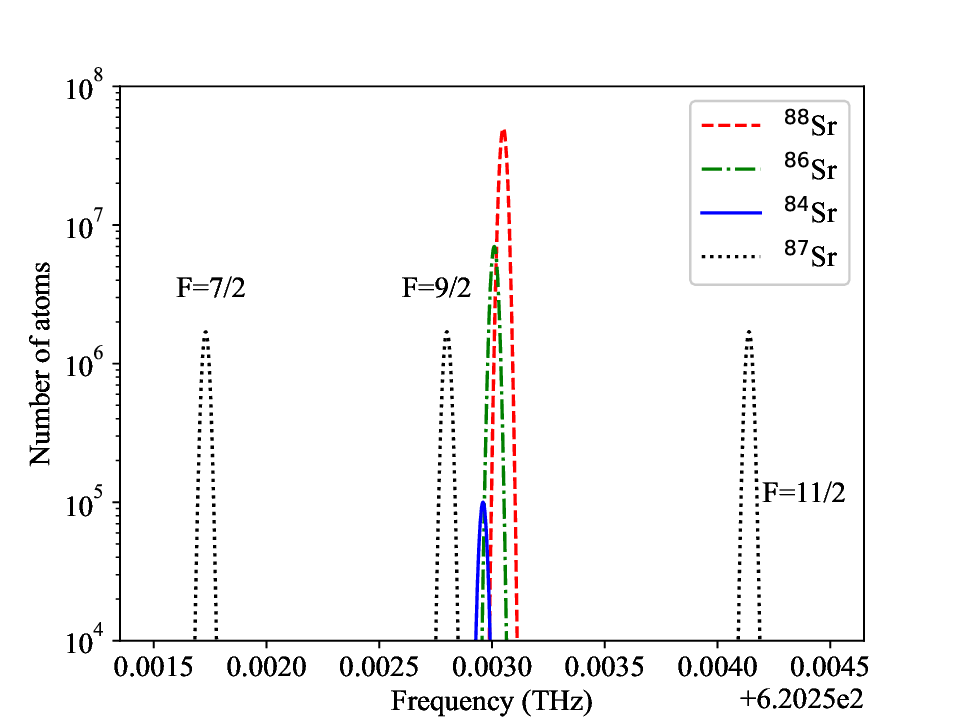}
	\caption[] { \label{fig:fig3} Spectroscopy data for the 
		$5s5p{\;^3}P_0 \rightarrow 5s5d{\;^3}D_1 $ transition obtained 
		from cold atoms in a magneto-optical trap using a fluorescence 
		detection method. The transitions for bosonic isotopes, $^{88}$Sr, $^{86}$Sr, 
		and $^{84}$Sr, are illustrated. To cover the hyperfine transition of 
		$^{87}$Sr, the laser is scanned across a 3 GHz range.}
\end{figure}

To perform the spectroscopy on the $5s5p{\;^3}P_0 \rightarrow 5s5d{\;^3}D_1$ transition for all the four stable isotopes, we continuously load the MOT and record the MOT fluorescence on the PMT. We load the targeted isotope in the MOT by changing the blue laser frequency to appropriate detuning. And then we apply 707 nm on-resonance light to repump $5s5p{\;^3}P_2$ state. The frequency of the 483 nm laser is then scanned resulting in an enhancement of the fluorescence from the MOT. Our data on MOT fluorescence as a function of 483 nm laser frequency is shown in Fig. \ref{fig:fig3}. As discernible from the figure, in the case of even isotopes, due to the absence of nuclear spin, the energy level structure is simple where we get one peak for each isotope. For the case of odd isotope, however, due to the coupling with nonzero nuclear spin ($I=9/2$), we observe three peaks at different frequencies, which is attributed to the splitting of $5s5d{\;^3}D_1$ state into three, $F = 7/2$, $9/2$, and $11/2$, hyperfine levels.

%%%%%%%%%%%%%%%%%%%%%%%%%%% figure

%%%%%%%%%%%%%%%%%%%%%%%%%%%%%%% table

\begin{center}
	\begin{table}[!h]
		\caption[]{\label{tab:tab2} Absolute transition frequencies of the 
			$5s5p$ $ ^3P_0 \rightarrow 5s5d$ $ ^3D_1 $ transition around 483 nm}
		\begin{tabular*}{\columnwidth}{@{\extracolsep{\stretch{1}}}*{4}{c}@{}} 	\hline 	\hline 
			{Isotope} & {$5s5p$ $ ^3P_0 \rightarrow 5s5d $ $^3D_1 $}  \\
			{} & {Absolute transition frequency (THz)}  \\
			\midrule
			88& 620.25305 (3)     \\
			86& 620.25301 (3)  \\
			84& 620.25296 (3)   \\
			87(F=7/2)& 620.25173 (3)   \\
			87(F=9/2)& 620.25280 (3)  \\
			87(F=11/2)& 620.25414 (3)   \\
			\hline 	\hline 
		\end{tabular*}
		\label{freq}
	\end{table}	
\end{center} 

  \begin{table*}
  	\centering
 	\caption{Magnetic dipole and electric quadrupole hyperfine structure 
 		constants for repumping state, $5s5d{\;^3}D_1$, and transition 
 		probabilities for transitions from $5s5d{\;^3}D_1$ to 
		$5s5p{^3}P_{0,1,2}$ and $5s5p{\;^1}P_1$ states in $^{87}$Sr.}

 		\begin{tabular}{cccccccc}\hline \hline
 			State/Transition & FSRCC & Breit & Self-energy & Vac.-Pol. & Breit + QED & Triples & Total\\
 			\hline     
 			\multicolumn{8}{c}{$A$ (in MHz)} \\ 
 			%	\cline{4-5}  \\
 			${^3D_1}$  &  239.89  &  0.03  &  -6.61  &  0.29  &  -6.29  &  6.96  
 			& 240.56   \\ \\
 			
 			\multicolumn{8}{c}{$B$ (in MHz)} \\ 
 			${^3D_1}$  &  5.98    & -0.01  &  0.35   &  0     &   0.34  & -0.54  
 			&  5.78      \\  \\
 			
 			\multicolumn{8}{c}{Transition Rate (in 10$^7$ s$^{-1}$)} \\ 
 			${^3P_0}\rightarrow {^3}D_1$ & 3.3337 & -0.0013  &  0.0649   &  0.0625  &  0.1262 & 0.0649  
 			& 3.5248  \\
 			${^3P_1}\rightarrow {^3D_1}$ & 3.1461 & -0.0002  &  0.0475  &  0.0470  &  0.0943 & 0.0476 
 			& 3.2879  \\
 			${^3P_2}\rightarrow {^3D_1}$ & 0.0780 & -0.0001  &  0.0018  &  0.0017  &  0.0034 & 0.0018 
 			& 0.0832 \\
 			${^1P_1}\rightarrow {^3D_1}$ & 0.0063 &  0.0 & -0.0004  & -0.0004  & -0.0007 & -0.0004 
 			& 0.0052 \\
 			\hline\hline
 		\end{tabular}

 	\label{prop_tab}
 \end{table*}

From our  measurement, we obtain the resonance frequency of the $5s5p\;{^3}P_0 \rightarrow 5s5d{\;^3}D_1$ transition in $^{88}$Sr as 620.25305(3) THz, which is a small shift ($\sim$13 GHz) and improvement over the previously available data of 620.2398(25) THz \cite{rubbmark}. The uncertainty in the transition frequency measurement arises mainly due to the limitation on the absolute accuracy of our wavemeter, which is 30 MHz. The absolute transition frequency for all the isotopes of strontium is listed in the Table \ref{freq}.

 \subsection{Hyperfine Structure Constants}

The hyperfine interaction in an atom provides a measure of the coupling between the nuclear electromagnetic moments and the electromagnetic fields of the electrons. Therefore, the determination of hyperfine splitting (HF) not only reveals insights into nuclear structure but also provides valuable information about the electronic wavefunction near the nucleus. 

In Table \ref{prop_tab}, we have provided the magnetic dipole, $A$, and electric quadrupole, $B$, hyperfine structure constants for the repumping state. For quantitative assessment of electron correlations, the contributions from the Breit and QED corrections are provided separately in the table. As evident from the table, self-energy has the significant contributions of $\approx$ 2.8 and 6.0\%, respectively, for $A$ and $B$. The contributions from the Breit interaction and vacuum polarization are, however, very small for both $A$ and $B$. The other significant contribution comes from the perturbative triples, which contributes $\approx$ 3.0 and 9.0\%, respectively,
to $A$ and $B$.

To compare with our measurements, using these hyperfine constants, we calculated the hyperfine splitting energies for all the allowed hyperfine levels. We compute splitting energies to be 1317, 241 and 1087 MHz, for $F = 7/2$, $9/2$ and $11/2$.

In Table \ref{freq}, we have compiled the absolute transition frequencies for $^{88}$Sr and $^{87}$Sr hyperfine levels, considering hyperfine levels $F = 7/2$, $9/2$, and $11/2$. This information enables the calculation of hyperfine splitting energies for each level in MHz, yielding values of 1320, 250, and 1090 MHz, respectively. Which are in excellent agreement with our calculated values. Comparing our theory results with the other available experiment, considering the complex nature of electron correlation in Sr, our value 240.6 MHZ for $A$ shows a good match with the result 227.3 MHz in Ref. \cite{stellmer}. To the best of our knowledge, however, there is no result from other theory calculations for comparison. Considering the hyperfine constant $B$, our calculated result is 5.78 MHz. This is in close agreement with our measured value 5.3 MHz. For comparing with other experiments, Ref. \cite{stellmer} reports an experimental value $0$, however, with a very large uncertainty of $\pm 10$ MHz. From other theory calculations, to the best of our search, we did not find any data for comparison.

%%%%%%%%%%%%%%%%%%%%%%%%%%%%%%% table
\subsection{Isotope Shift}
Next, to assess the effect of nuclear mass and charge distributions, we calculated the isotope shifts in the transition frequency of ${^1}S_0 \rightarrow {^3}P_1$ and ${^3}P_0 \rightarrow {^3}D_1$ transitions for $^{84, 86, 87}$Sr relative to $^{88}$Sr. For this, the ASFs were calculated using the MCDHF theory using GRASP18 code \cite{fischer} and then shifts in the frequencies were calculated using the RIS4 program \cite{ekman}. We started with \{$5s4d, 5s5d$\} and \{$5s5p, 4d5p$\} multireference even and odd parity valence configurations, respectively, which embeds the ASFs of our interests -- ${^1}S_0$, ${^3}P_1$, ${^3}P_0$ and ${^3}D_{1}$ -- and then the electron correlations were added in the subsequent calculations. To include the electron correlations from the {\em valence-to-virtual}, {\em core-to-valence} and {\em core-to-virtual} excitations, we performed a series of MCDHF calculations where we added virtual orbitals to active space layer wise. For each correlation layer, the single and double excitations of electrons were considered from the valence and core orbitals. The maximum correlation layer used in the active spaces of even and odd configurations were up to \{9s9p7d7f\} and \{9s9p7d5f\}, respectively. It is should, however, be mentioned that, for {\em core-to-valence} and {\em core-to-virtual} excitations, orbitals up to $3d$ were considered as frozen core. We restricted the active apace to this to find a close match with the experimental excitation energies. The maximum error achieved in the excitation energies of ${^1}S_0 \rightarrow {^3}P_1$ and ${^3}P_0 \rightarrow {^3}D_1$ transitions are 6\% and 2\%, respectively.

%%%%%%%%%  Table: Isotope Shift     %%%
\begin{table*}
	\centering
	\caption{Nuclear rms radius, $\delta \langle r^{2} \rangle$ (fm$^{2}$), 
		reduced mass (a.m.u.), isotope shift (MHz), and modified isotope 
		shift (MHz a.m.u.), for ${^1}S_0 \rightarrow {^3}P_1$ and ${^3}P_0 \rightarrow {^3}D_1$ 
		transitions for $^{84,86,87}$Sr relative to $^{88}$Sr.} 
		\begin{tabular}{ccccccc} \hline\hline
			Isotope (A)  & $\delta \langle r^{2} \rangle$ &  $\mu_{A}$  & 
			\multicolumn{2}{c}{${^1}S_{0}$ -  ${^3}P_1$}  & 
			\multicolumn{2}{c}{$^3P_{0}$ - ${^3}D_{1}$}  \\  
			\cline{4-5}  \cline{6-7}
			& &  & $\delta \nu_{A}$ & $\mu \delta \nu_{A}$ 
			& $\delta \nu_{A}$ & $\mu \delta \nu_{A}$ \\
			\hline    
			$^{84}$Sr & 0.487473 & 1848 & 346.57 & 640350.48 & 85.28 & 157597.44 \\
			$^{86}$Sr & 0.242665 & 3784 & 167.06 & 632155.04 & 41.12 & 155598.08  \\
			$^{87}$Sr & 0.121069 & 7656 &  82.03 & 628021.68 & 20.19 & 154574.64   \\
			\hline\hline
		\end{tabular}
	\label{iso_tab}
\end{table*}	   	   
%%%%%%%%%%%%%

In Table \ref{iso_tab}, we have provided our calculated results on isotope shifts, $\delta \nu_A$, and modified isotope shifts, $\mu \delta \nu_A$, for the above discussed transitions. The values of nuclear {\em rms} radius and reduced mass, $\mu = AA'/A-A'$, used in the calculation of $\delta \nu_A$ and $\mu \delta \nu_A$ are also given. Considering the ${^1}S_0 \rightarrow {^3}P_1$ transition, our calculated $\delta \nu_A$ for $^{84}$Sr, $^{86}$Sr and $^{87}$Sr are 346.56, 167.06 and 82.03 MHz, respectively. Except for $^{87}$Sr, these are in excellent agreement with the experimental values 351.49, 163.82 and 62.19, respectively, reported in Ref. \cite{miyake}. The reason for the large discrepancy in the case of $^{87}$Sr could be attributed to the different nature of electron correlations in this fermionic isotope and need to be taken care separately. After optimizing it separately, we obtained an isotope shift of 61 MHz with an active space configurations up to $7s7p7d5f$. The error in the excitation energy is, however, large, about 13\%. Here, we would like to mention an important trend we observed from a series of MCDHF calculations to extract isotope shifts. We find that, when wavefunctions are optimized more accurately with the inclusion of additional layers of electron correlations, to improve the accuracy of excitation energies, it leads to deteriorated isotope shifts with respect to experiments. The reason for this could be attributed to the inherent shortcomings associated with the treatment of electron correlations in MCDHF theory. For the isotope shift for $^3P_0 \rightarrow ^3D_1$ transition, we get 85.28, 41.12 and 20.19 MHz for $^{84}$Sr, $^{86}$Sr and $^{87}$Sr, respectively. 

In the experiment, where laser cooling was applied to different isotopes and spectroscopy was performed, as detailed in the methodology section, we obtained the absolute frequencies for each isotope. These absolute frequency measurements are presented in Table \ref{freq}. Using this data, we calculated the isotope shifts for each isotope, resulting in corresponding values of 90, 40, and 32 MHz, for $^{84}$Sr, $^{86}$Sr and $^{87}$Sr, respectively, and these are in good match with our calculated values. To the best of our knowledge, there are no results available on isotope shifts from other theory calculations for comparison for this transition.

% 1.  6%, 13612 cm^-1, 9s9p6d even, 9s9p5d odd
% 13%, 19325, 6s6p6d even, 6ss6p5d odd
% exp, 14504
% 13%, 12504 cm^-1, 61 MHz, 7s7p7d5f even, 7s7p6d5f odd 
%1S0 - 3P0 13418, 6%; 394 MHz 84; 195, 86; 97 MHz, 87 --> exp. 349, 162, 62; 9s9p6d even, 9s9p5d odd,  
% 438, 214, 106, ee 21315, exp 20689, 3% --> 9s9p6d even, 9s9p5d odd
% 174, 85, 41, ee 21116, exp 20689, 2% -->9s9p9d7f even, 9s9p9d5f odd
%9s9p9d7f even, 9s9p9d5f odd --> 1S0 - 3P1, 12580; 235, 118, 59.5 MHz

%%%%%%%%%%%%%%%%%%%   Conclusion  %%%%%%%%%%%%%%%%%%%%%%%%%%%%%%
\section{Conclusion}
In conclusion, we conducted spectroscopy on the $5s5p$ $ ^3 P_0 \rightarrow 5s5d$ $ ^3 D_1 $ transition for all stable isotopes of Sr, determining the absolute transition frequency with an accuracy of 30 MHz. We introduced the utilization of the $5s5p$$ ^3 P_0 \rightarrow 5s5d$ $ ^3 D_1 $ transition of Sr for repumping $5s5p{\;^3}P_0 $ atoms in strontium, comparing it with the conventionally employed 679 nm laser scheme. Our results illustrate the successful repumping of all stable isotopes of Sr using the 483 nm laser. Fock-space relativistic coupled-cluster calculations were performed to calculate the branching ratio, linewidth, hyperfine splitting constant, and isotope shift for the $5s5p$ $ ^3 P_0 \rightarrow 5s5d$ $ ^3 D_1 $ transition, and our experimental measurements of hyperfine splitting constants and isotope shift are in good agreement with the calculated values.
%%%%%%%%%%%%%%%%%%%%%%%  Acknowledgment %%%%%%%%%%%%%%%%%%%%%%%%%%%%%%

\section{Acknowledgments}
The authors would like to thank the Department of
Science and Technology, Govt. of India and the Council
for Scientific and Industrial Research (CSIR, Govt. of India) for research funding.
KP and KB would like to thank IISER Pune for institutional
fellowships. SSM and PD would like to thank Council for Scientific
and Industrial Research research fellowships. The authors would like to express their gratitude and acknowledge Jay Mangaonkar for their fruitful and valuable discussions. Palki, BKM and UDR would like to acknowledge 
Prof. Dilip Angom for their fruitful and valuable discussions. The authors would also 
like to acknowledge the funding support from I-HUB Quantum Technology Foundation 
through the National Mission on Interdisciplinary Cyber-Physical Systems (NM-ICPS) 
of the Department of Science and Technology, Govt. of India.
We would also like to thank the support of the IIT Delhi, New Delhi for 
providing the High Performance Computing cluster, Padum.

\section*{Disclosures}
The authors declare no conflicts of interest.

\section*{References}
\bibliography{Ref}
\clearpage

\end{document}